# On the Structure and Efficient Computation of IsoRank Node Similarities

Ehsan Kazemi and Matthias Grossglauser

School of Computer and Communication Sciences, EPFL
*E-mail address:* `firstname.lastname@epfl.ch`

**Abstract**

The alignment of protein-protein interaction (PPI) networks has many applications, such as the detection of conserved biological network motifs, the prediction of protein interactions, and the reconstruction of phylogenetic trees [1, 2, 3]. IsoRank is one of the first global network alignment algorithms [4, 5, 6], where the goal is to match all (or most) of the nodes of two PPI networks. The IsoRank algorithm first computes a pairwise node similarity metric, and then generates a matching between the two node sets based on this metric. The metric is a convex combination of a structural similarity score (with weight $\alpha$) and an extraneous amino-acid sequence similarity score for two proteins (with weight $1 - \alpha$).

In this short paper, we make two contributions. First, we show that when IsoRank similarity depends only on network structure ($\alpha = 1$), the similarity of two nodes is only a function of their degrees. In other words, IsoRank similarity is invariant to any network rewiring that does not affect the node degrees. This result suggests a reason for the poor performance of IsoRank in structure-only ($\alpha = 1$) alignment.

Second, using ideas from [7, 8], we develop an approximation algorithm that outperforms IsoRank (including recent versions with better scaling, e.g., [9]) by several orders of magnitude in time and memory complexity, despite only a negligible loss in precision.

## 1 Introduction

We first define the IsoRank algorithm as given in [5]. Assume we are given two networks $G_1(V_1, E_1)$ and $G_2(V_2, E_2)$ with $|V_i| = n_i$ and $|E_i| = m_i$. Let $N_{i,u}$ represent the neighbours of node $u$ in graph $i$ and $d_{i,u} = |N_{i,u}|$ is its degree. Also, assume $\boldsymbol{b}$ is the doubly indexed vector of BLAST sequence similarities of proteins, i.e., $\boldsymbol{b}[u, v]$ is the sequence similarity between proteins $u \in V_1$ and $v \in V_2$. The vector $\boldsymbol{e} = \frac{\boldsymbol{b}}{|\boldsymbol{b}|_1}$ is the normalized vector of sequence similarity scores. Also, $\boldsymbol{P}$ is a $n_1 n_2 \times n_1 n_2$ square matrix, where $\boldsymbol{P}[u_1, u_2][v_1, v_2]$ refers to the entry at row $(u_1, u_2)$ and column $(v_1, v_2)$.[1] The elements of $\boldsymbol{P}$ are defined

---
[1]Both the rows and columns are doubly indexed.

as follows:

$$\boldsymbol{P}[u_1, u_2][v_1, v_2] = \begin{cases} \frac{1}{d_{1,v_1} d_{2,v_2}}, & \text{if } (u_1, v_1) \in E_1 \text{ and } (u_2, v_2) \in E_2. \\ 0, & \text{otherwise} \end{cases}$$

**Problem 1** (IsoRank similarity problem [5])**.** *Find the vector $\boldsymbol{r}$ from*

$$\boldsymbol{r} = \alpha \boldsymbol{P} \boldsymbol{r} + (1 - \alpha) \boldsymbol{e}, \qquad (1)$$

*for $0 \leq \alpha \leq 1$. If we assume $|\boldsymbol{r}|_1 = 1$ then the problem is equivalent to finding $\boldsymbol{r}$ from*

$$\boldsymbol{r} = \left( \alpha \boldsymbol{P} + (1 - \alpha) \boldsymbol{e} \mathbb{1}^T \right) \boldsymbol{r}.$$

The first step of the IsoRank algorithm is to compute $\boldsymbol{r}$, where $\boldsymbol{r}[u, v]$ corresponds to the similarity between nodes $u \in G_1$ and $v \in G_2$. The value of $\boldsymbol{r}[u, v]$ can be interpreted as a likelihood such that the node $u$ aligns to the node $v$ based on structural and sequence similarities. The second step is to construct an alignment based on the similarity vector $\boldsymbol{r}$. The original paper proposes two approaches for alignment: (i) solving the maximum-weight bipartite graph matching, where edge weights are elements of $\boldsymbol{r}$; (ii) greedy alignment, which matches the most similar nodes first and removes them, then matches the most similar among the remaining, etc. [4, 5]. While the greedy method is much faster, it has shown slightly better alignment quality in many cases [4].

## 2 Structural IsoRank Only Depends on Degrees

In this section, we show that structure-only IsoRank ($\alpha = 1$) depends on the edge sets $E_{1,2}$ of the two graphs only through node degrees. This is a surprisingly weak dependence on the network structure, in the sense that any rewiring that conserves node degrees does not affect the alignment produced by IsoRank.

We first define the tensor product (Kronecker product) of two graphs.

**Definition 1** (Tensor product of two graphs [10])**.** *The tensor product $G_1 \times G_2$ of two graphs $G_1(V_1, E_1)$ and $G_2(V_2, E_2)$ is the graph $G(V, E)$ defined as follows:*

- *$V = V_1 \times V_2$ is the Cartesian product of the two sets $V_{1,2}$.*
- *There is an edge between $(u_1, u_2)$ and $(v_1, v_2) \in V$ (i.e., $((u_1, u_2), (v_1, v_2)) \in E$) if and only if $(u_1, v_1) \in E_1$ and $(u_2, v_2) \in E_2$.*

**Lemma 1.** *The IsoRank similarity problem is equivalent to the PageRank problem [11] over the graph $G = G_1 \times G_2$ with the teleportation constant $\alpha$ and the preference vector $\boldsymbol{e}$.*

*Proof.* Call $\boldsymbol{A}$ the adjacency matrix of graph $G$. Based on the definition of tensor product of two graphs, it is easy to show that the degree of node $(v_1, v_2) \in V$ is $d_{1,v_1} d_{2,v_2}$. Then the PageRank problem over graph $G$ with preference vector $\boldsymbol{e}$ is to find $\boldsymbol{r}'$, such that

$$\boldsymbol{r}' = \alpha \boldsymbol{D}^{-1} \boldsymbol{A} \boldsymbol{r}' + (1 - \alpha) \boldsymbol{e}, \qquad (2)$$

where $\boldsymbol{D}$ is the diagonal matrix of weighted degrees. Again, it is straightforward to see that $\boldsymbol{P} = \boldsymbol{D}^{-1} \boldsymbol{A}$. From these two facts, we conclude that $\boldsymbol{r} = \boldsymbol{r}'$. □

**Lemma 2.** *For the case $0 < \alpha < 1$ we have*

$$\boldsymbol{r} = (1 - \alpha)(\boldsymbol{I} - \alpha\boldsymbol{P})^{-1}\boldsymbol{e}.$$

*Proof.* The equation is derived simply from (1). We need only show that $\boldsymbol{I} - \alpha\boldsymbol{P}$ is non-singular. To prove this, note that $\boldsymbol{I} - \alpha\boldsymbol{P}$ is a strictly diagonally dominant matrix. From the Levy-Desplanques theorem [12], we know that a strictly diagonally dominant matrix is non-singular. □

Next, we explain how to compute $\boldsymbol{r}$ efficiently for three different cases (i) $\alpha = 0$ (ii) $\alpha = 1$ and (iii) $0 < \alpha < 1$.

For the case $\alpha = 0$, the trivial answer is $\boldsymbol{r} = \boldsymbol{e}$.

**Lemma 3.** *For the IsoRank similarity problem with $\alpha = 1$, we have*

$$\boldsymbol{r} = \left[\frac{d_{1,1}d_{2,1}}{m}, \cdots, \frac{d_{1,u}d_{2,v}}{m}, \cdots, \frac{d_{1,n_1}d_{2,n_2}}{m}\right] \quad (3)$$

*where $m = \sum_{i \in V_1} \sum_{j \in V_2} d_{1,i}d_{2,j}$.*

*Proof.* In this case, the IsoRank similarity problem is equivalent to the PageRank problem over the undirected graph $G$ with $\boldsymbol{e} = 0$. It is easy to show that the vector $\boldsymbol{r}$ is the steady state probability distribution of a random walk over $G$. It is a well-known result that this probability distribution is proportional to the degree of each node [13]. The lemma follows from the fact that the degree of a node $(v_1, v_2) \in V$ is $d_{1,v_1}d_{2,v_2}$ and the elements of $\boldsymbol{r}$ should sum to one. □

From Lemma 3, we conclude that when IsoRank uses only the structural properties of the two input graphs $G_{1,2}$ (i.e., $\alpha = 1$), the similarity of two nodes $u \in G_1$ and $v \in G_2$ is only a function of their degrees $d_{1,u}$ and $d_{2,v}$. This means that in this case IsoRank matches nodes only based on the product of their degrees. In particular, the matching generated by the greedy approach is as follows: (i) the node with highest degree in $G_1$ is matched to the highest degree node in $G_2$; (ii) then the unmatched nodes with the second highest degrees in the two graphs are matched, and so on.

**Example 1.** *We illustrate this insight using an example taken from [5], see Figure 1. The equations are for the case where $\alpha = 1$. The goal of the IsoRank similarity problem is to find the values of $\boldsymbol{r}_{ij}$. It is easy to see that the product of the degrees of the nodes (as stated by Lemma 3) is the non-trivial answer for this set of equations, e.g., if we have $\boldsymbol{r}_{bb'} = 2 \times 2 = 4, \boldsymbol{r}_{ac'} = 3, \boldsymbol{r}_{ca'} = 3, \boldsymbol{r}_{aa'} = 1$ and $\boldsymbol{r}_{cc'} = 9$, then*

$$\boldsymbol{r}_{bb'} = \frac{1}{3}\boldsymbol{r}_{ac'} + \frac{1}{3}\boldsymbol{r}_{ca'} + \boldsymbol{r}_{aa'} + \frac{1}{9}\boldsymbol{r}_{cc'}.$$

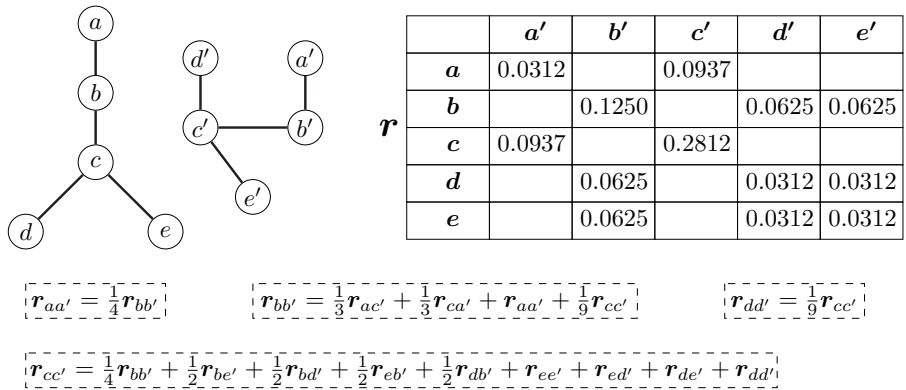

$$r_{aa'} = \tfrac{1}{4} r_{bb'}$$ $$r_{bb'} = \tfrac{1}{3} r_{ac'} + \tfrac{1}{3} r_{ca'} + r_{aa'} + \tfrac{1}{9} r_{cc'}$$ $$r_{dd'} = \tfrac{1}{9} r_{cc'}$$

$$r_{cc'} = \tfrac{1}{4} r_{bb'} + \tfrac{1}{2} r_{be'} + \tfrac{1}{2} r_{bd'} + \tfrac{1}{2} r_{eb'} + \tfrac{1}{2} r_{db'} + r_{ee'} + r_{ed'} + r_{de'} + r_{dd'}$$

Figure 1: A small example from [5] of the IsoRank similarity problem for the case $\alpha = 1$.

We performed experiments to confirm the invariance of IsoRank to degree-conserving rewirings in the case $\alpha = 1$. We generated two correlated graphs $G_{1,2}$ using the random bigraph model from [14, 15], and compute the IsoRank similarity vector with $\alpha = 1$. We randomly rewire some edges from both $G_{1,2}$ such that node degrees are preserved (using the method from [16]), and then align the two rewired graphs using the IsoRank implementation of the GraphM package [17].[2] This experiment confirmed Lemma 3. In conclusion, we observe that output of IsoRank with $\alpha = 1$ is only a function of node degrees and is otherwise independent of graph structure.

**Corollary 1.** *For $n = \max(|V_1|, |V_2|)$ and $\alpha \in \{0, 1\}$, we can compute the IsoRank vector in $n^2$ steps. Also, we can compute the similarity between any two nodes in $O(1)$.*

Note that in (3) the normalizing constant $m = \sum_{i \in V_1} \sum_{j \in V_2} d_{1,i} d_{2,j} = \sum_{i \in V_1} d_{1,i} \sum_{j \in V_2} d_{2,j} = 4|E_1||E_2|$ can be computed in $O(1)$ assuming the total sizes of the edge sets are available.

**Corollary 2.** *For $n = \max(|V_1|, |V_2|)$ and $\alpha = 1$, we can find the output of greedy IsoRank in $O(n \log n)$ steps.*

Corollary 2 is because we can separately order the nodes in $G_1$ and $G_2$ by decreasing degree, then match the two lists.

## 3 Fast Approximate IsoRank

For the case $0 < \alpha < 1$, we can use the results of [7, 8] to approximate node-pair similarities efficiently. From the result of Lemma 1, we know that the IsoRank similarity problem is equivalent to the PageRank problem over an undirected graph. The authors of [8] designed an approximate algorithm for solving the PageRank problem over undirected graphs with tight error bounds. Their algorithm is an improved version of the algorithm from [7].

---
[2]To the best of our knowledge, this package is currently the most faithful implementation of the IsoRank algorithm, which is why we used it for this experiment.

Assume $e[u,v]$ is the normalized sequence similarity between two nodes $u \in G_1$ and $v \in G_2$. The `SharpApproximateIsoRank` algorithms returns $\tilde{r}$ as the approximation of $r$, where $\tilde{r}[u,v]$ is the (approximate) total similarity between $u$ and $v$. Algorithm 1 describes `SharpApproximateIsoRank`.[3]

---

**Algorithm 1:** SharpApproximateIsoRank($e, \alpha, \epsilon$)

1. $\beta \leftarrow \frac{1-\alpha}{\alpha}, \epsilon' \leftarrow 1, \tilde{e} \leftarrow e$ and $\tilde{r} \leftarrow 0$ ;
2. **while** $\epsilon' > \epsilon$ **do**
3. $\quad \epsilon' \leftarrow \epsilon'/2$;
4. $\quad \tilde{r}', \tilde{e}' \leftarrow$ ApproximateIsoRank($\tilde{e}, \beta, \epsilon'$) ;
5. $\quad \tilde{r} \leftarrow \tilde{r} + \tilde{r}'$ and $\tilde{e} \leftarrow \tilde{e}'$ ;
6. **end**

---

**Algorithm 2:** ApproximateIsoRank($e, \beta, \epsilon$)

1. $\tilde{r} \leftarrow 0$ and $\tilde{e} \leftarrow e$ ;
2. **while** *there exists at least a pair $(u,v) \in V$ such that $\tilde{e}[u,v] \geq \epsilon d_{1,u} d_{2,v}$* **do**
3. $\quad$ pick any pair $(u,v) \in V$ such that $\tilde{e}[u,v] \geq \epsilon d_{1,u} d_{2,v}$;
4. $\quad \tilde{r}, \tilde{e} \leftarrow$ Push$((u,v), \tilde{r}, \tilde{e})$;
5. **end**
6. **return** $\tilde{r}$ *and* $\tilde{e}$;

---

**Algorithm 3:** Push$((u,v), \tilde{r}, \tilde{e}, \beta)$

1. $\tilde{r}' \leftarrow \tilde{r}$ and $\tilde{e}' \leftarrow \tilde{e}$ ;
2. $\tilde{r}'[u,v] \leftarrow \tilde{r}[u,v] + \frac{\beta}{2+\beta}\tilde{e}[u,v]$ ;
3. $\tilde{e}'[u,v] \leftarrow \frac{1}{2+\beta}\tilde{e}[u,v]$ ;
4. **for** *each pair $(u',v')$ such that $(u',u) \in E_1$ and $(v',v) \in E_2$* **do**
5. $\quad \tilde{e}'[u',v'] \leftarrow \tilde{e}[u',v'] + \frac{\tilde{e}[u,v]}{(2+\beta)d_{1,u}d_{2,v}}$ ;
6. **end**
7. **return** $\tilde{r}'$ *and* $\tilde{e}'$;

---

**Lemma 4.** *For the number of edges $m$ in the product graph $G$, we have $m \leq \min(2|E_1|D_2, 2|E_2|D_1)$, where $D_{1,2}$ are the maximum degrees in the two networks.*

*Proof.* From $m = \sum_{i \in V_1} \sum_{j \in V_2} d_{1,i} d_{2,j}$ we conclude that $m \leq \sum_{i \in V_1} d_{1,i} D_2 \leq 2|E_1|D_2$ and $m \leq 2|E_2|D_1$. □

**Theorem 1.** *The* `SharpApproximateIsoRank` *algorithm reruns the two vectors $\tilde{r}$ and $\tilde{e}$ such that*
$$\tilde{r} = \alpha P \tilde{r} + (1-\alpha)(e - \tilde{e}),$$

---

[3] The same as `SharpApproximatePR` from [7].

where $|\frac{\tilde{e}(u,v)}{d_{1,u}d_{2,v}}| \leq \epsilon$ for all pairs $(u,v) \in V$. The running time of the algorithm is

$$O\left(\frac{(1+\alpha)\min(|E_1|D_2, |E_2|D_1)\log(1/\epsilon)}{1-\alpha}\right).$$

*Proof.* This theorem is the direct result of Theorem 2 from [8] and Lemmas 1 and 4. □

**Corollary 3.** *For $n = \max(|V_1|, |V_2|)$ and given constants $c > 0, 0 < \alpha < 1$, we can approximate the IsoRank vector $r$ in $O(n^3 \log(n))$ steps with $\epsilon = \Omega(n^{-c})$.*

Note that the time complexity of the original IsoRank algorithm for computing the approximate IsoRank vector is $O(n^4)$. Corollary 3 gives the worst case performance. For many real (sparse) biological networks, time complexity is much smaller.

## 4 Simulation Result

We compared the performance of the original IsoRank algorithm[4] with our implementation of the `SharpApproximateIsoRank` algorithm on aligning PPI networks of the five major eukaryotic species. Table 1 provides a brief description of the PPI networks that are extracted from the IntAct database [18, 19]. The amino-acid sequences of proteins for each species are collected in the FASTA format from the UniProt database [20, 21]. The BLAST bit-score similarities [22] are calculated using these amino acid sequences. The IsoRank algorithm[5] took 13 hours and 31 minutes to perform all ten pairwise alignments between species from Table 1.[6] The `SharpApproximateIsoRank` algorithm performed these ten alignments in 53 minutes for $\epsilon = 10^{-12}$, 59 minutes for $\epsilon = 10^{-13}$, and one hour and 11 minutes for $\epsilon = 10^{-14}$. For larger networks, the relative advantage of `SharpApproximateIsoRank` would be even more pronounced.

Table 1: PPI networks of five major eukaryotic species from the IntAct molecular interaction database [18, 19].

| species | #nodes | #edges | Avg. deg. |
|---|---|---|---|
| C. elegans | 4950 | 11550 | 4.67 |
| D. melanogaster | 8532 | 26289 | 6.16 |
| H. sapiens | 19141 | 83312 | 8.71 |
| M. musculus | 10765 | 22345 | 4.15 |
| S. cerevisiae | 6283 | 76497 | 24.35 |

---

[4]The official IsoRank implementation from `http://groups.csail.mit.edu/cb/mna/isobase/`

[5]Run with parameters `--K 50 --thresh 1e-5 --alpha 0.9 --maxveclen 1000000`. Note that in this version of IsoRank the parameter `--maxveclen` sets a limit on the number of non-zero entries in the IsoRank vector, e.g., $10^6$ out of $\approx 2 \times 10^8$ possible entries between H. sapiens and M. musculus in this example.

[6]The GraphM package [17] took several days to finish these alignments.

# 5  Conclusion

We have shown that the IsoRank node similarity metric has a peculiar structure, in that the network (structural) similarity depends only on the nodes' degrees, and not on the actual edge set of the two networks. It appears that this fact has not been noted before, and provides some insight into its relatively poor performance for $\alpha = 1$. We have also shed light on the relationship between the IsoRank and PageRank problems. The IsoRank similarity problem is in fact equivalent to applying PageRank over the Kronecker product of the two graphs. This equivalence enables us to apply ideas for efficient PageRank approximation algorithms to the IsoRank similarity problem, with significant gains in runtime and memory complexity.

**Acknowledgement**  We would like to thank Hamed Hassani for our early discussions on this problem.

# References

[1] Oleksii Kuchaiev and Nataša Pržulj. Integrative network alignment reveals large regions of global network similarity in yeast and human. *Bioinformatics*, 27(10):1390–1396, 2011.

[2] Ahed Elmsallati, Connor Clark, and Jugal Kalita. Global alignment of protein-protein interaction networks: A survey. *Computational Biology and Bioinformatics, IEEE/ACM Transactions on*, PP(99):1–1, 2015.

[3] Fazle E Faisal, Lei Meng, Joseph Crawford, and Tijana Milenković. The post-genomic era of biological network alignment. *EURASIP Journal on Bioinformatics and Systems Biology*, 2015(1):1–19, 2015.

[4] Rohit Singh, Jinbo Xu, and Bonnie Berger. Pairwise Global Alignment of Protein Interaction Networks by Matching Neighborhood Topology. In *Proc. of Research in Computational Molecular Biology 2007*, San Francisco, CA, USA, April 2007.

[5] Rohit Singh, Jinbo Xu, and Bonnie Berger. Global alignment of multiple protein interaction networks with application to functional orthology detection. *Proceedings of the National Academy of Sciences*, 105(35):12763–12768, 2008.

[6] Chung-Shou Liao, Kanghao Lu, Michael Baym, Rohit Singh, and Bonnie Berger. IsoRankN: spectral methods for global alignment of multiple protein networks. *Bioinformatics*, 25(12):i253–i258, 2009.

[7] Reid Andersen, Fan Chung, and Kevin Lang. Local Graph Partitioning using PageRank Vectors. In *Proc. of 47th Annual IEEE Foundations of Computer Science (FOCS) 2006*, Berkeley, CA, USA, October 2006.

[8] Fan Chung Graham and Wenbo Zhao. A Sharp PageRank Algorithm with Applications to Edge Ranking and Graph Sparsification. In *Proc. of 7th International Workshop on Algorithms and Models for the Web-Graph (WAW) 2010*, Stanford, CA, USA, December 2010.

[9] Mohsen Bayati, David F Gleich, Amin Saberi, and Ying Wang. Message Passing Algorithms for Sparse Network Alignment. *ACM Transactions on Knowledge Discovery from Data (TKDD)*, 7(1):3, 2013.

[10] Richard Hammack, Wilfried Imrich, and Sandi Klavžar. *Handbook of product graphs*. CRC press, 2011.

[11] Amy N Langville and Carl D Meyer. *Google's PageRank and Beyond: The Science of Search Engine Rankings*. Princeton University Press, 2011.

[12] Roger A Horn and Charles R Johnson. *Matrix Analysis*. Cambridge university press, 2012.

[13] László Lovász. Random walks on graphs: A survey. *Combinatorics, Paul erdos is eighty*, 2(1):1–46, 1993.

[14] Ehsan Kazemi, Lyudmila Yartseva, and Matthias Grossglauser. When Can Two Unlabeled Networks Be Aligned Under Partial Overlap? In *Proc. of the 53rd Annual Allerton Conference on Communication, Control, and Computing*, 2015.

[15] Ehsan Kazemi, S. Hamed Hassani, and Matthias Grossglauser. Growing a Graph Matching from a Handful of Seeds. *Proc. of the VLDB Endowment*, 8(10):1010–1021, 2015.

[16] Sergei Maslov and Kim Sneppen. Specificity and stability in topology of protein networks. *Science*, 296(5569):910–913, 2002.

[17] GraphM. Graph Matching package. http://cbio.ensmp.fr/graphm/.

[18] Henning Hermjakob, Luisa Montecchi-Palazzi, Chris Lewington, Sugath Mudali, Samuel Kerrien, Sandra Orchard, Martin Vingron, Bernd Roechert, Peter Roepstorff, Alfonso Valencia, et al. IntAct: an open source molecular interaction database. *Nucleic Acids Research*, 32(suppl 1):D452–D455, 2004.

[19] IntAct. IntAct: an open source molecular interaction database. http://www.ebi.ac.uk/intact/, 2014.

[20] Rolf Apweiler, Amos Bairoch, Cathy H Wu, Brigitte Barker, Winona Cand Boeckmann, Serenella Ferro, Elisabeth Gasteiger, Hongzhan Huang, Rodrigo Lopez, Michele Magrane, et al. UniProt: the universal protein knowledgebase. *Nucleic Acids Research*, 32(suppl 1):D115–D119, 2004.

[21] UniProt. UniProt: the universal protein knowledgebase. http://www.uniprot.org/, 2014.

[22] Stephen F Altschul, Warren Gish, Webb Miller, Eugene W Myers, and David J Lipman. Basic local alignment search tool. *Journal of Molecular Biology*, 215(3):403–410, 1990.